\documentclass[sigconf,table,10pt]{acmart}
\AtBeginDocument{%
  }

\setcopyright{acmlicensed}
\copyrightyear{2025}
\acmYear{2025}
\acmDOI{XXXXXXX.XXXXXXX}
\acmConference[Conference acronym 'XX]{Make sure to enter the correct
  conference title from your rights confirmation emai}{June 03--05,
  2025}{Woodstock, NY}
\acmISBN{978-1-4503-XXXX-X/18/06}

\usepackage{lipsum}

\usepackage{algorithm}
\usepackage{algpseudocode}

\usepackage{url}

\begin{document}

\title{Distributed Retrieval-Augmented Generation}


\author{Chenhao Xu}
\email{chenhao.xu@vu.edu.au}
\orcid{0000-0003-0819-7269}
\affiliation{%
  \institution{Victoria University}
  \city{Melbourne}
  \country{Australia}
}

\author{Longxiang Gao}
\email{gaolx@sdas.org}
\orcid{0000-0002-3026-7537}
\affiliation{%
  \institution{Qilu University of Technology}
  \city{Jinan}
  \country{China}
}

\author{Yuan Miao}
\email{yuan.miao@vu.edu.au}
\orcid{0000-0002-6712-3465}
\affiliation{%
  \institution{Victoria University}
  \city{Melbourne}
  \country{Australia}
}

\author{Xi Zheng}
\email{james.zheng@mq.edu.au}
\orcid{0000-0002-2572-2355}
\affiliation{%
  \institution{Macquarie University}
  \city{Sydney}
  \country{Australia}
}

\renewcommand{\shortauthors}{XXX et al.}

\begin{abstract}
As large language models (LLMs) become increasingly adopted on edge devices, Retrieval-Augmented Generation (RAG) is gaining prominence as a solution to address factual deficiencies and hallucinations by integrating external knowledge.
However, centralized RAG architectures face significant challenges in data privacy and scalability. 
For instance, smart healthcare services often rely on collecting sensitive patient data and building a centralized knowledge base to provide better diagnosis and treatment advice, while privacy concerns significantly impede this process.
Besides, maintaining a comprehensive and continuously updated knowledge base is costly, particularly in response to regional epidemics and rapidly mutating viruses.
To address these challenges, this paper introduces Distributed Retrieval-Augmented Generation (DRAG), a novel framework that improves data privacy by eliminating the need for a centralized knowledge base and restoring data control to owners.
DRAG incorporates a Topic-Aware Random Walk (TARW) algorithm that leverages LLMs to extract query topics and facilitate targeted peer discovery within a peer-to-peer network, enabling efficient knowledge retrieval in decentralized environments.
Extensive experiments across three diverse datasets and LLMs demonstrate that DRAG with TARW achieves near-centralized RAG performance by using half as many messages as flooding.
The code is available at \url{https://github.com/xuchenhao001/DRAG}.
\end{abstract}

\ccsdesc[500]{Computing methodologies~Natural language processing}
\ccsdesc[500]{Computing methodologies~Distributed artificial intelligence}

\keywords{Large Language Model, Retrieval-Augmented Generation, Distributed Computing, Edge Computing}

\received{20 February 2007}
\received[revised]{12 March 2009}
\received[accepted]{5 June 2009}

\maketitle

\section{Introduction}
\label{sec:introduction}

Large Language Models (LLMs) have revolutionized natural language processing, enabling a wide range of applications in edge computing~\cite{zheng2024review}. From voice assistants like Siri and Google Assistant to personalized recommendation systems, LLMs are integral to enhancing user experience on edge devices~\cite{qu2025mobile}. In healthcare, LLMs are being explored for symptom checking and medical advice~\cite{wang2023pre}. In driver assistance, LLMs can improve navigation and real-time traffic updates~\cite{zhou2024vision}. However, a significant challenge with LLMs is their propensity to generate factually incorrect information, a phenomenon known as hallucination~\cite{ji2023survey}, which hampers reliability in critical applications.

To mitigate this, Retrieval Augmented Generation (RAG)~\cite{gao2023retrieval} has emerged as a promising approach. RAG combines LLM generative capabilities with external knowledge sources, retrieving relevant information before generating responses to ensure factual grounding. In edge computing, RAG can be particularly beneficial for providing up-to-date and accurate information directly to users on their devices~\cite{qu2025mobile}.

However, centralized RAG architectures present considerable hurdles, particularly concerning data privacy and scalability~\cite{fan2024survey, zyskind2023don,koga2024privacy, muhamed2024cache, zeng2024mitigating, purwar2024evaluating}. 
Consider a smart healthcare application: while leveraging LLMs to provide personalized treatment recommendations, these systems often require access to sensitive patient data to build and maintain a comprehensive knowledge base. However, privacy concerns and regulations usually restrict the sharing of such protected health information. 
Similarly, in autonomous driving systems, accessing the locations and traffic plans of other vehicles can improve navigation efficiency and safety. However, aggregating such data into a centralized server raises significant privacy concerns.

In addition, the huge amount of data generated by regional epidemics, rapidly mutating viruses, and vehicle movements pose centralized RAG a scalability challenge, requiring efficient management of the computational demands for both retrieval and generation. 
Beyond these technical challenges, maintaining a continuously updated and comprehensive centralized knowledge base is inherently costly, requiring substantial resources for data acquisition, storage, and processing.

To address the aforementioned challenges, this paper introduces Distributed Retrieval-Augmented Generation (DRAG), a novel framework that improves data privacy and scalability of RAG. DRAG eliminates the need for a central knowledge base or direct data sharing by enabling peer-to-peer (P2P) knowledge retrieval and generation across a distributed network. The decentralized design mitigates privacy risks by allowing users to have full control over their data while distributing the computational burden of knowledge base maintenance and retrieval. 
To facilitate efficient knowledge discovery within the DRAG network, a novel Topic-Aware Random Walk (TARW) algorithm is proposed. TARW leverages LLMs to extract query topics and guide peer discovery while utilizing a local cache to further enhance performance.
The key contributions of this paper are summarized as follows:
\begin{itemize}
    \item A decentralized RAG framework tailored for edge computing environments, improving data privacy and scalability by eliminating direct data sharing and distributing the knowledge retrieval burden among peers.
    \item A novel Topic-Aware Random Walk (TARW) algorithm that leverages LLMs to extract query topics and guide efficient peer discovery within the DRAG network through a local cache.
    \item Extensive experiments across diverse datasets and LLMs validate that DRAG with TARW achieves performance comparable to centralized RAG with significantly reduced communication overhead than flooding.
\end{itemize}

The rest of the paper is organized as follows: Section~\ref{sec:related_work} reviews related work on Retrieval-Augmented Generation (RAG) scalability, privacy, and security. Section~\ref{sec:model} details the proposed DRAG framework and the TARW algorithm. Section~\ref{sec:experiments} presents the experimental evaluation. Finally, Section~\ref{sec:conclusion} concludes the paper.

\section{Related Work}
\label{sec:related_work}

This section reviews related work on DRAG, focusing on RAG scalability, privacy, and security.

\subsection{RAG Scalability}

Retrieval-Augmented Generation (RAG) has emerged as a pivotal approach to mitigate hallucinations in LLMs~\cite{chen2024benchmarking}. Lewis et al.~\cite{lewis2020retrieval} pioneered this field with their seminal work introducing RAG as a framework that combines dense retrieval with sequence-to-sequence models for knowledge-intensive NLP tasks. Despite recent advancements in retrieval mechanisms~\cite{guu2020retrieval,karpukhin2020dense,asai2023self,edge2024local,hu2024grag,mavromatis2024gnn,yu2025rankrag}, centralized RAG architectures inherently face challenges regarding computational scalability and knowledge base maintenance.

Some researchers have proposed query routing~\cite{shnitzer2024large,mu2025unsupervised} and hierarchical retrieval~\cite{zhang2024hierarchical,wang2025archrag} methods that primarily focus on efficiently assigning queries or aggregating results across multiple LLM candidates. While these approaches mitigate RAG scalability issues to some extent, they still overlook the fundamental scalability challenges associated with centralized knowledge base storage and processing within each LLM candidate.

Recent research has also explored federated RAG architectures~\cite{wang2024feb4rag, addison2024c}, which enable federated search and collaborative knowledge base development. However, their approaches still rely on a centralized request mediator, which inherently introduces vulnerability points and scalability bottlenecks.

\subsection{RAG Privacy and Security}

Privacy and security concerns within RAG systems have become increasingly critical as these technologies are deployed in sensitive domains~\cite{zeng2024good}. For instance, studies have demonstrated the feasibility of privacy attacks that can extract confidential data from the RAG knowledge base~\cite{jiang2024rag}. Research has also revealed vulnerabilities of RAG to membership inference attacks~\cite{anderson2024my,liu2024mask,li2024generating} and poisoning attacks~\cite{xue2024badrag}.

Consequently, various privacy-preserving techniques have been explored in the context of RAG, including secure multi-party computation~\cite{zyskind2023don}, homomorphic encryption~\cite{zhao2024frag}, watermarking~\cite{lv2025rag}, and differential privacy~\cite{grislain2024rag,koga2024privacy,zeng2024mitigating}. While these advancements represent significant progress, they typically rely on centralized knowledge bases or processing nodes. This inherent reliance introduces potential points of vulnerability and can compromise data sovereignty.

By contrast, the DRAG framework proposed in this paper addresses these limitations by empowering users with control over their data. DRAG allows individuals to determine what information is shared within the network and what remains private, thereby ensuring data privacy.

\section{Proposed Model}
\label{sec:model}

This section explains the proposed model, covering the Distributed Retrieval-Augmented Generation framework, the Topic-Aware Random Walk algorithm, and further discussion.

\subsection{Distributed RAG}

\begin{figure*}[htbp]
    \centering
    \includegraphics[width=\linewidth]{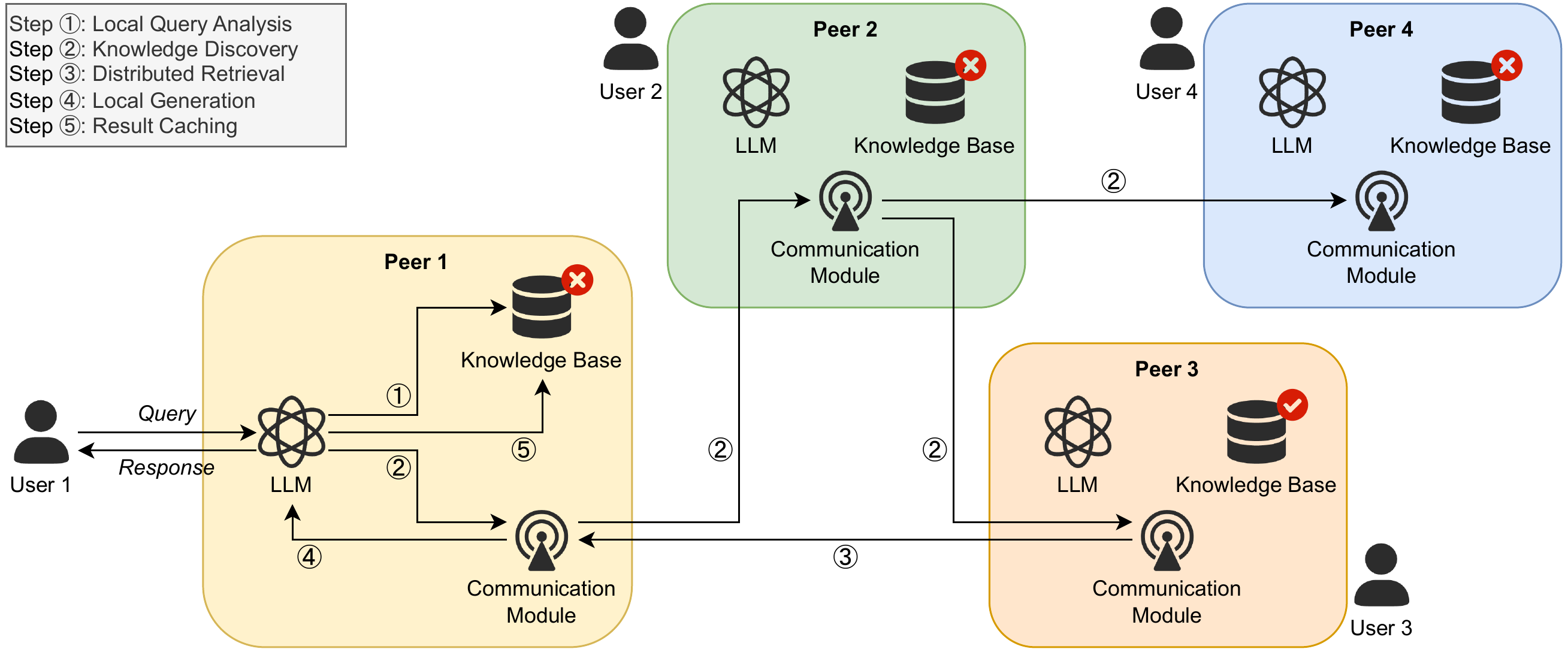}
    \caption{Overview of the Distributed RAG (DRAG) Architecture. DRAG employs a peer-to-peer network where each peer maintains a local knowledge base and utilizes the Topic-Aware Random Walk (TARW) algorithm for distributed knowledge retrieval. The query process involves local query analysis, knowledge discovery, distributed retrieval, local generation, and result caching.}
    \label{fig:drag_architecture}
\end{figure*}

Distributed RAG represents a paradigm shift from traditional centralized RAG systems. Unlike conventional approaches that rely on a centralized knowledge base, DRAG distributes both the knowledge and computation across a network of edge devices, creating a P2P framework for knowledge sharing and query resolution. This architecture enables knowledge sharing and collaborative query resolution, potentially improving scalability, resilience, and privacy compared to centralized approaches.

Figure~\ref{fig:drag_architecture} illustrates the architecture of DRAG. Each edge device is referred to as a peer. In healthcare, peers could be smartphones, wearable devices, provider workstations, or hospital servers sharing anonymized insights. In autonomous driving, peers could be individual vehicles, roadside units, or smart cameras exchanging traffic information. Each peer is equipped with a local knowledge base, a local LLM instance, and a communication module facilitating direct peer-to-peer interaction. Upon receiving a query, a peer (querying peer) first attempts to answer it using its local knowledge base. If the query cannot be adequately answered locally, the querying peer initiates a distributed search across the network, leveraging the TARW algorithm. Once receiving relevant knowledge from other peers, the querying peer uses LLM to aggregate and generate a final answer.

When a user issues a query, DRAG processes it through the following steps:

\begin{enumerate}
    \item \textbf{Local Query Analysis}: The query processing flow in DRAG begins with a local query analysis phase. The local LLM on the querying peer analyzes the query to extract key topics and determine if external knowledge is required. This determination can be made through several methods. One approach involves calculating the semantic similarity between the query and the content in the local knowledge base. If the highest similarity score falls below a predefined threshold, external knowledge is deemed necessary. Alternatively, the system can incorporate user feedback signals, where explicit or implicit signals, such as low satisfaction scores on initial local responses, trigger the need for external knowledge.
    \item \textbf{Knowledge Discovery}: If external information is deemed necessary, the system enters the knowledge discovery phase, initiating the TARW algorithm to discover peers possessing relevant knowledge. The TARW algorithm intelligently explores the network, prioritizing peers that have demonstrated expertise in the topic areas identified during the query analysis phase.
    \item \textbf{Distributed Retrieval}: Following the discovery of relevant peers, DRAG proceeds to the distributed retrieval phase. Rather than retrieving raw data, DRAG gathers user-filtered knowledge snippets from peers to ensure data privacy. In particular, these filters can be rule-based, relying on predefined patterns and keywords, or LLM-driven, utilizing machine learning models to automatically identify and mask sensitive data. By integrating these privacy-preserving mechanisms, DRAG facilitates secure and efficient knowledge sharing while minimizing the risks of data leakage and privacy violations.
    \item \textbf{Local Generation}: At this stage, the querying peer combines the retrieved knowledge snippets with its local context to generate an accurate and coherent response using its local LLM. 
    \item \textbf{Result Caching}: Finally, query results, consisting of the generated response and the source knowledge snippets, are cached locally to improve response time for future similar queries. Besides, the system also caches information about the expertise areas of neighboring peers. This can be achieved by storing topics of their successful contributions to previous queries. This cached expertise information enables the system to intelligently route future queries directly to the peers most likely to possess relevant knowledge, further reducing network traffic and improving overall efficiency.
\end{enumerate}

The DRAG architecture offers several advantages over centralized RAG systems. By keeping sensitive data on local devices, DRAG enhances privacy, making it particularly suitable for domains like healthcare where data confidentiality is crucial. The decentralized approach also improves scalability by distributing the computational load across multiple nodes, reducing bottlenecks associated with a centralized knowledge base. Additionally, result caching minimizes the latency of peer discovery and query processing, ensuring faster responses while reducing redundant computations. Finally, the distributed nature of this framework increases robustness against single points of failure, ensuring uninterrupted service even if some nodes become unavailable.

\subsection{Topic-Aware Random Walk}

\begin{algorithm}
\caption{Topic-Aware Random Walk (TARW)}
\label{alg:tarw}
\begin{flushleft}
\textbf{Input:} Query $q$, Max hops $H_{max}$, Neighbor selection count $k$, Relevance threshold $\theta$. \\
\textbf{Output:} Privacy-filtered knowledge $\mathcal{K}$ or $\emptyset$.
\end{flushleft}
\begin{algorithmic}[1]
    \State $\mathcal{T} \gets \text{ExtractTopic}(q)$ \Comment{Use local LLM}
    \State $\mathcal{V} \gets \{p_0\}$ \Comment{$p_0$ is the querying peer}
    \State $\mathcal{Q} \gets \{(p_0, 0)\}$ \Comment{Queue of (peer, hop count) pairs}
    \State $\mathcal{E} \gets \emptyset$ \Comment{Cache of peer expertise}

    \While{$\mathcal{Q} \neq \emptyset$}
        \State $(p_i, h) \gets \text{Dequeue}(\mathcal{Q})$
        \If{$h \geq H_{max}$}
            \State \textbf{continue}
        \EndIf
        
        \State $\mathcal{K}_i \gets \text{QueryLocal}(p_i, q)$
        \If{$\text{Relevance}(\mathcal{K}_i, q) \geq \theta$}
            \State $\mathcal{E} \gets \mathcal{E} \cup \{(p_i, \mathcal{T})\}$
            \State \Return \text{PrivacyFilter}($\mathcal{K}_i$)
        \EndIf
        
        \State $\mathcal{N}_i \gets \text{Neighbors}(p_i) \setminus \mathcal{V}$
        \State $\mathcal{S} \gets \text{RelevanceScore}(\mathcal{N}_i, \mathcal{E}, \mathcal{T})$
        \State $\mathcal{P}_{next} \gets \text{SelectTopK}(\mathcal{N}_i, \mathcal{S}, k)$
        
        \For{$p_j \in \mathcal{P}_{next}$}
            \State $\mathcal{V} \gets \mathcal{V} \cup \{p_j\}$
            \State $\text{Enqueue}(\mathcal{Q}, (p_j, h + 1))$
        \EndFor
    \EndWhile
    \State \Return $\emptyset$
\end{algorithmic}
\end{algorithm}

The knowledge discovery phase in DRAG leverages a novel routing algorithm called TARW, designed to efficiently navigate the distributed network and identify peers possessing relevant knowledge. Unlike traditional random walk algorithms, which explore the network randomly, TARW incorporates topic awareness to prioritize the exploration of peers more likely to hold the desired information. The algorithm is outlined in Algorithm~\ref{alg:tarw}.

TARW begins by extracting key topics $\mathcal{T}$ from the input query $q$ using the local LLM of the querying peer $p_0$ (Line 1). This process provides a semantic understanding of the information being sought. A set of visited peers $\mathcal{V}$ and a queue $\mathcal{Q}$ are initialized, with the querying peer and a hop count of 0 added (Lines 2-3). Additionally, a set $\mathcal{E}$ is initialized to cache peer expertise (Line 4). The algorithm then iterates while the queue is not empty (Line 5).

In each iteration, the algorithm dequeues a peer $p_i$ and its associated hop count $h$ from the queue (Line 6). If the hop count exceeds a predefined maximum number of hops, $H_{max}$, the algorithm skips to the next iteration to prevent excessively long paths (Lines 7-8). The peer then attempts to answer the query locally by consulting its local knowledge base (Line 10). If the relevance score of the locally retrieved knowledge $\mathcal{K}_i$ to the query $q$, as determined by a relevance function (e.g., semantic similarity), exceeds a predefined threshold $\theta$, the algorithm adds the peer $p_i$ and the extracted topics $\mathcal{T}$ to the set of cached peer expertise $\mathcal{E}$ (Line 12). This dynamically updates the system's knowledge of peer specializations. The algorithm then applies a privacy filter to the retrieved knowledge and returns the filtered result (Line 13). This ensures that only relevant and privacy-preserving knowledge is shared.

If the local query does not yield sufficiently relevant information, the algorithm identifies the neighbors $\mathcal{N}_i$ of the current peer $p_i$ that have not yet been visited (Line 15). It then calculates relevance scores $\mathcal{S}$ for all neighbors based on their historical expertise $\mathcal{E}$ and the extracted topics $\mathcal{T}$ (Line 16). This allows the algorithm to prioritize neighbors that have previously demonstrated expertise in the topics relevant to the current query. The algorithm selects the top $k$ neighbors ($\mathcal{P}_{next}$) with the highest relevance scores (Line 17). These neighbors are then added to the visited set $\mathcal{V}$ and enqueued in $\mathcal{Q}$ with an incremented hop count (Lines 18-21).

If the algorithm exhausts all possible paths within the defined hop limit without finding sufficiently relevant knowledge, it returns an empty set (Line 23). The parameters $H_{max}$ and $k$ are crucial for balancing the exploration and exploitation of the network, with $H_{max}$ controlling the search depth and $k$ determining the breadth of the search at each hop. By dynamically caching peer expertise and using this information to guide the random walk, TARW significantly improves the efficiency and effectiveness of knowledge discovery in DRAG, while enabling the system to adapt to the evolving knowledge landscape of the distributed network.

\subsection{Further Discussion}

Beyond the core mechanics of DRAG and the TARW algorithm, several critical aspects that may impact the feasibility of the proposed model are discussed in this section.

\textbf{Privacy:} While DRAG inherently addresses some privacy concerns by eliminating the centralized knowledge base, it introduces new challenges that need careful consideration. First, the user-filtered knowledge snippets shared between peers may still inadvertently leak sensitive information. Advanced techniques, such as differential privacy~\cite{grislain2024rag,koga2024privacy}, could be integrated to further anonymize the shared knowledge and protect against potential inference attacks. Second, the TARW algorithm itself could be vulnerable to deanonymization attacks if an adversary can observe the network traffic and infer the query topics and the expertise of individual peers. Defenses against such attacks might involve adding noise to the topic extraction process, employing mix networks to obfuscate the routing paths, or implementing access control mechanisms to restrict peer visibility. While this risk exists in general, deploying DRAG within a controlled environment, such as a consortium network of hospitals and clinics, can offer a layer of enhanced privacy. In such scenarios, all participating parties sign privacy-preserving agreements that establish clear guidelines for data sharing and usage, reducing the risk of malicious actors within the network. Furthermore, the defined roles and responsibilities within the consortium allow for better auditing and accountability, facilitating the detection and prevention of privacy breaches. 

\textbf{Scalability:} DRAG offers promising scalability benefits compared to centralized RAG due to its distributed architecture. The computational load is distributed across multiple peers, avoiding the bottleneck of a central server. However, the scalability of DRAG depends on the efficiency of the TARW algorithm and the connectivity of the P2P network. In densely connected networks, TARW can quickly discover relevant peers. However, in sparsely connected networks, the algorithm may require more hops to reach the desired knowledge, increasing latency and network traffic. Employing network clustering techniques or implementing super-peers can help improve connectivity and reduce the average path length. Fault tolerance is another key advantage of DRAG. Since the knowledge is distributed across multiple peers, the system can continue to function even if some peers become unavailable. The robustness of DRAG is further enhanced by the result caching mechanism, where knowledge snippets are replicated across multiple peers after queries to provide redundancy. The choice of caching strategy depends on the trade-off between storage overhead and fault tolerance requirements.

\textbf{Knowledge Reliability:} A major challenge in DRAG is ensuring the reliability and consistency of the shared knowledge. Unlike centralized knowledge bases that can be carefully curated and maintained, the knowledge in DRAG is distributed across multiple peers with varying levels of expertise and data quality. This can lead to issues such as misinformation, stale knowledge, and biased information. To address these challenges, DRAG can incorporate mechanisms for knowledge validation and ranking. Peers can rate the quality and relevance of the knowledge snippets they receive, and this feedback can be used to weigh the contributions of different peers. Additionally, consensus mechanisms can be employed to ensure that conflicting knowledge snippets are resolved in a consistent manner. The trustworthiness of individual peers is also an important factor. Reputation systems can be used to track the historical performance of peers and reward those who consistently provide high-quality knowledge.

\textbf{Incentive Mechanisms:} The success of DRAG depends on the willingness of peers to contribute their knowledge and computational resources to the network. However, peers may be reluctant to share their resources if they do not receive adequate compensation. Therefore, incentive mechanisms are needed to encourage participation and ensure the long-term sustainability of the system~\cite{xu2021lightweight}. Several incentive mechanisms could be employed in DRAG. One approach is to use a token-based system, where peers are rewarded with tokens for contributing knowledge, answering queries, and validating information. These tokens can then be exchanged for services within the network, such as priority access to information or increased computational resources. Another approach is to use a reputation-based system, where peers with high reputations are given preferential treatment. This can incentivize peers to maintain high-quality knowledge and provide accurate responses. Finally, a knowledge marketplace can be established, where peers can sell their knowledge to other peers in exchange for payment. This can create a financial incentive for peers to contribute valuable knowledge to the network. The design of effective incentive mechanisms is a complex task that requires careful consideration of the trade-offs between participation, fairness, and efficiency.

\section{Experiments}
\label{sec:experiments}

This section presents the experimental evaluation, covering the experimental setup, performance comparison, and sensitivity analysis.

\subsection{Experimental Setup}

This section details the experimental setup used to evaluate the performance of DRAG. DRAG is compared against two baseline systems: CRAG (Centralized RAG) and NoRAG (LLM only, without RAG). Within the DRAG framework, the effectiveness of the proposed TARW algorithm is evaluated against two classic resource search algorithms in P2P networks: Random Walk (RW) and Flooding (FL).

\textbf{Network Configuration:}  Peer-to-peer networks are simulated using the NetworkX\footnote{\url{https://networkx.org}} library in Python. Specifically, the Barab{\'a}si-Albert model~\cite{albert2000error} is utilized to mimic real-world network topologies like the Internet or social media. Network sizes are explored with varying numbers of peers, specifically $20$, $40$, $60$, $80$, and $100$. The default number of peers was set to $20$ unless otherwise specified. The connectivity of the network, representing the number of peers each peer is directly connected to, was also varied across values of $2$, $4$, $6$, and $8$, with a default connectivity of $4$.

\textbf{Algorithm Parameters:} The TARW algorithm's performance is influenced by several parameters. The maximum number of hops ($H_{max}$) is fixed to $6$. Different values for the neighbor selection count ($k$), which determines the number of neighbors considered for forwarding the query in each hop, are explored, specifically using values of $2$, $4$, $6$, and $8$, with a default value of $4$. The relevance threshold ($\theta$), which dictates the minimum relevance score required for a knowledge snippet to be considered relevant, is set to $0.8$.

\textbf{Large Language Models:} To evaluate the generalizability of DRAG, three open-source LLMs of comparable size are employed: Llama 3.2-3B\footnote{\url{https://ollama.com/library/llama3.2:3b}} (by default), Gemma 2 2B\footnote{\url{https://ollama.com/library/gemma2:2b}}, and Qwen2.5 3B\footnote{\url{https://ollama.com/library/qwen2.5:3b}}. This allows for assessing the performance of DRAG across different model architectures and training datasets.

\textbf{Datasets:} The performance of DRAG is evaluated using three diverse datasets, each presenting unique challenges:
\begin{itemize}
    \item \textbf{MMLU\footnote{\url{https://huggingface.co/datasets/cais/mmlu}}:} A multiple-choice question answering benchmark designed to evaluate general knowledge of LLM across a wide range of subjects.
    \item \textbf{Medical Extended\footnote{\url{https://huggingface.co/datasets/sarus-tech/medical_extended}}:} A synthetic dataset comprising generated patient questions and corresponding doctor answers on fabricated symptoms, diseases, and treatments. This dataset simulates a healthcare knowledge domain where expertise is paramount.
    \item \textbf{News\footnote{\url{https://huggingface.co/datasets/heegyu/news-category-dataset}}:} A dataset containing approximately 210,000 news headlines from HuffPost, spanning the years 2012 to 2022. In the experiments, the LLM is tasked with identifying the author of a news article based solely on its headline, requiring the retrieval of relevant contextual information.
\end{itemize}

\textbf{Evaluation Metrics:} The following metrics are used to evaluate the performance of RAG:
\begin{itemize}
    \item \textbf{\# Hops:} The average number of hops taken by the query to retrieve relevant knowledge. This measures the efficiency of the knowledge retrieval process in the distributed environment.
    \item \textbf{\# Messages:} The average number of messages sent during the knowledge retrieval process.  This is a metric for evaluating the communication overhead of DRAG.
    \item \textbf{Hit Rate:} The percentage of queries for which relevant knowledge was successfully retrieved.  This indicates the overall effectiveness of the knowledge retrieval process.
    \item \textbf{EM (Exact Match):} The percentage of predicted tokens that exactly match their corresponding ground truth tokens, providing a strict measure of accuracy.
    \item \textbf{F1 Score:} The harmonic mean of precision and recall, providing a balanced measure of LLM-generated output.
    \item \textbf{Precision:} The ratio of correctly predicted tokens to the total number of predicted tokens, measuring the quality of the generated output.
    \item \textbf{Recall:} The ratio of correctly predicted tokens to the total number of ground truth tokens, measuring the completeness of the generated output.
\end{itemize}

\subsection{Performance Comparison}

\begin{table*}[htbp]
\caption{Comparative Performance Analysis of DRAG Variants with Centralized and Non-Augmented Baselines.}
\label{table:drag_baseline}
\centering
\begin{tabular}{llllllllll}
\toprule
Scheme & LLM & Dataset & \# Hops & \# Messages & Hit Rate & EM & F1 & Precision & Recall \\
\midrule
NoRAG     & Llama 3.2-3B & MMLU    & -    & -     & -    & 17.30\% & 17.41\% & 17.38\% & 17.51\% \\
CRAG      & Llama 3.2-3B & MMLU    & -    & -     & 99.81\% & 85.73\% & 85.75\% & 85.74\% & 85.76\% \\
DRAG-RW   & Llama 3.2-3B & MMLU    & 5.13 & 5.36  & 23.46\% & 19.97\% & 19.97\% & 19.97\% & 19.97\% \\
DRAG-FL   & Llama 3.2-3B & MMLU    & 1.68 & 10.91 & 100.00\% & 85.43\% & 85.45\% & 85.44\% & 85.46\% \\
\rowcolor{gray!50}
DRAG-TARW & Llama 3.2-3B & MMLU    & 1.72 & 6.87  & 98.11\% & 83.90\% & 83.92\% & 83.92\% & 83.93\% \\
\hline
NoRAG     & Llama 3.2-3B & Medical & -    & -     & -    & 0.00\%  & 32.28\% & 29.39\% & 40.56\% \\
CRAG      & Llama 3.2-3B & Medical & -    & -     & 99.81\% & 82.80\% & 94.46\% & 99.49\% & 93.43\% \\
DRAG-RW   & Llama 3.2-3B & Medical & 4.94 & 5.22  & 27.62\% & 19.66\% & 24.32\% & 25.81\% & 24.02\% \\
DRAG-FL   & Llama 3.2-3B & Medical & 1.62 & 9.72  & 99.86\% & 77.78\% & 91.58\% & 96.56\% & 90.54\% \\
\rowcolor{gray!50}
DRAG-TARW & Llama 3.2-3B & Medical & 1.88 & 8.82  & 98.67\% & 77.07\% & 90.58\% & 95.66\% & 89.51\% \\
\hline
NoRAG     & Llama 3.2-3B & News    & -    & -     & -    & 0.08\%  & 0.67\%  & 0.42\%  & 3.25\%  \\
CRAG      & Llama 3.2-3B & News    & -    & -     & 99.16\% & 69.07\% & 76.66\% & 80.08\% & 75.82\% \\
DRAG-RW   & Llama 3.2-3B & News    & 5.16 & 5.38  & 21.96\% & 28.97\% & 16.83\% & 17.56\% & 16.65\% \\
DRAG-FL   & Llama 3.2-3B & News    & 1.74 & 10.99 & 99.20\% & 68.95\% & 76.54\% & 79.95\% & 75.69\% \\
\rowcolor{gray!50}
DRAG-TARW & Llama 3.2-3B & News    & 1.88 & 7.82  & 96.86\% & 67.62\% & 74.73\% & 78.18\% & 73.86\% \\
\bottomrule
\end{tabular}
\end{table*}

\begin{figure}
    \centering
    \includegraphics[width=\linewidth]{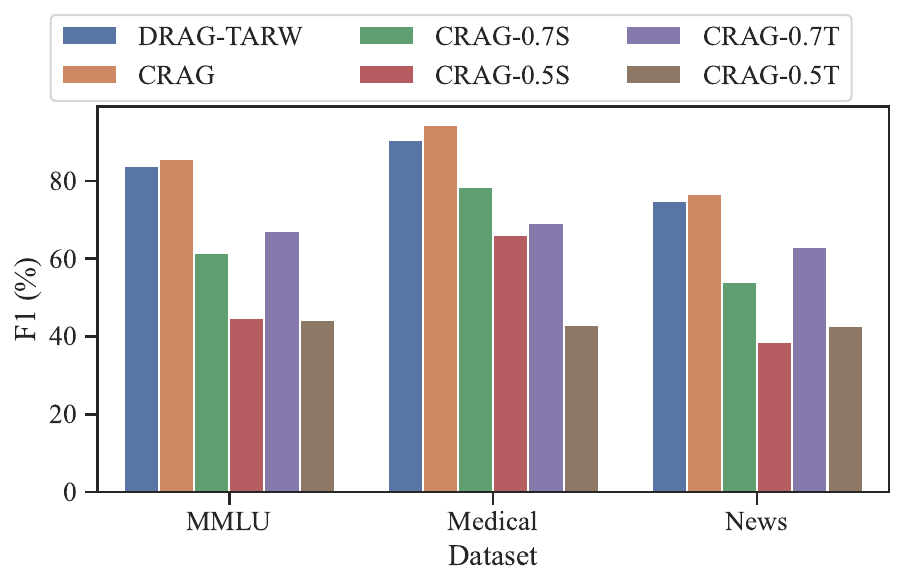}
    \caption{Comparative F1 Scores of DRAG and Centralized RAG Variants Under Varying Knowledge Base Completeness. CRAG-0.7S means the centralized knowledge base contains $70\%$ of knowledge snippets, while CRAG-0.7T means it includes $70\%$ of topics.}
    \label{fig:f1_scheme_datasets}
\end{figure}

\begin{figure*}[htpb]
    \centering
    \includegraphics[width=\linewidth]{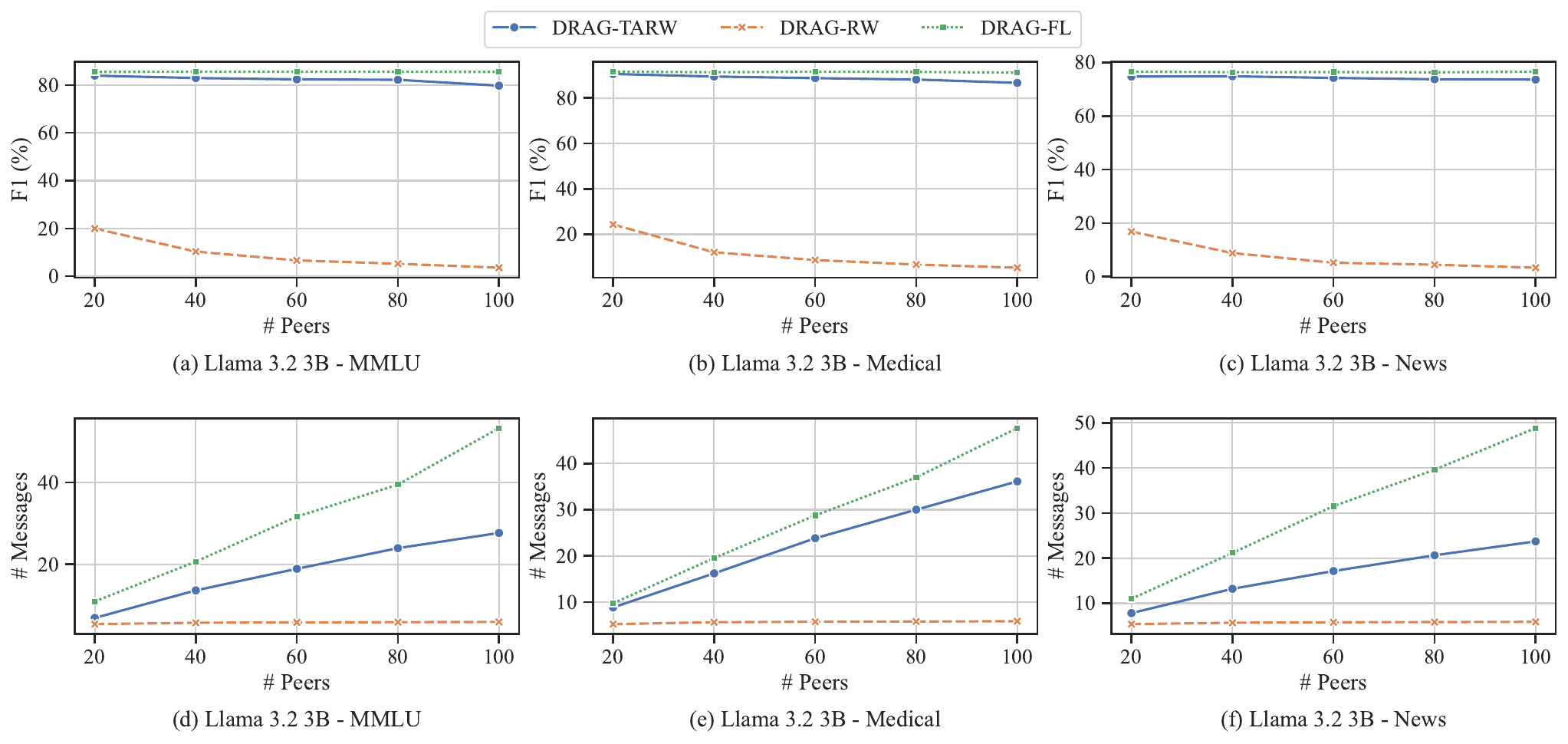}
    \caption{Impact of network size on F1 score (first row) and message overhead (second row) in DRAG.}
    \label{fig:f1_anm_peer}
\end{figure*}

\begin{figure}
    \centering
    \includegraphics[width=\linewidth]{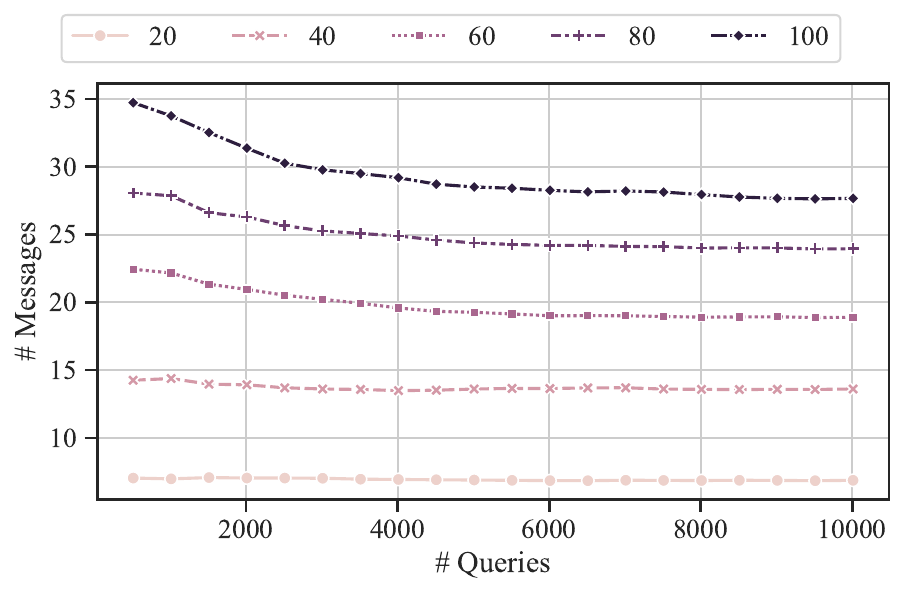}
    \caption{Convergence of average message counts with increasing query number in DRAG-TARW. Each line represents a different network size, with $20$, $40$, $60$, $80$, and $100$ indicating the number of peers in the network.}
    \label{fig:anm_nq}
\end{figure}

\begin{figure}
    \centering
    \includegraphics[width=\linewidth]{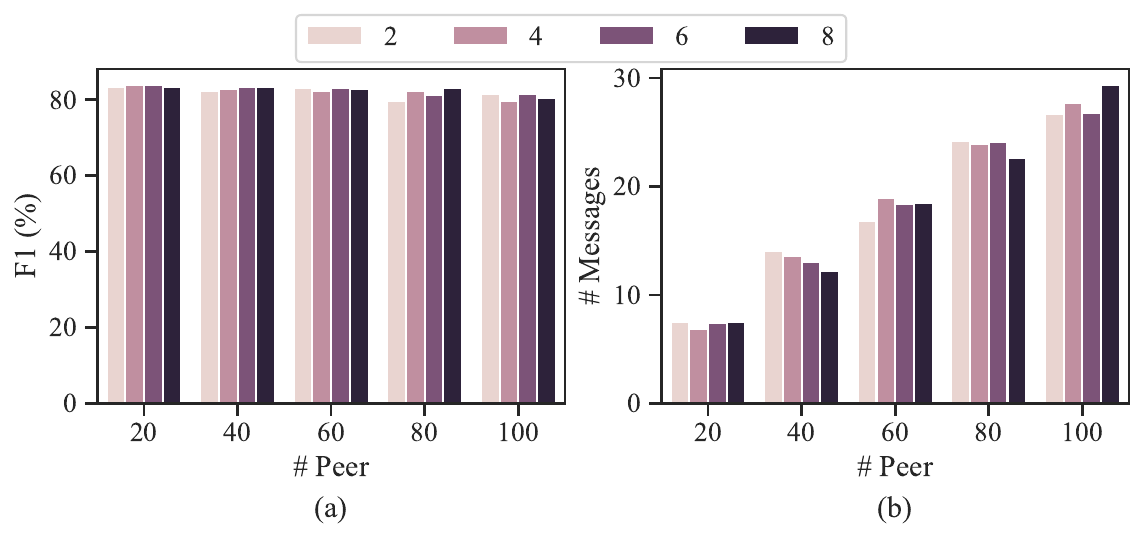}
    \caption{Impact of peer connectivity on performance (a) and communication cost (b) in DRAG-TARW. Peer connectivity is defined as the number of edges attached from a new node to existing nodes ($2$, $4$, $6$, and $8$), consistent with the Barab{\'a}si-Albert model.}
    \label{fig:sensitivity_peer_attach}
\end{figure}

\begin{figure}
    \centering
    \includegraphics[width=\linewidth]{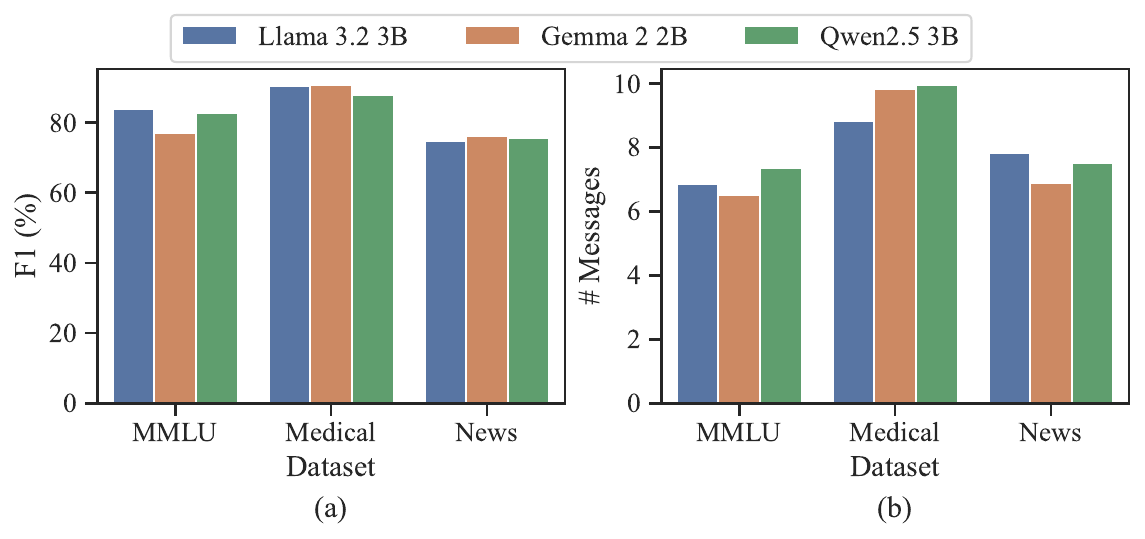}
    \caption{Influence of large language model on DRAG-TARW performance (a) and communication cost (b).}
    \label{fig:sensitivity_llm_datasets}
\end{figure}

\begin{figure}
    \centering
    \includegraphics[width=\linewidth]{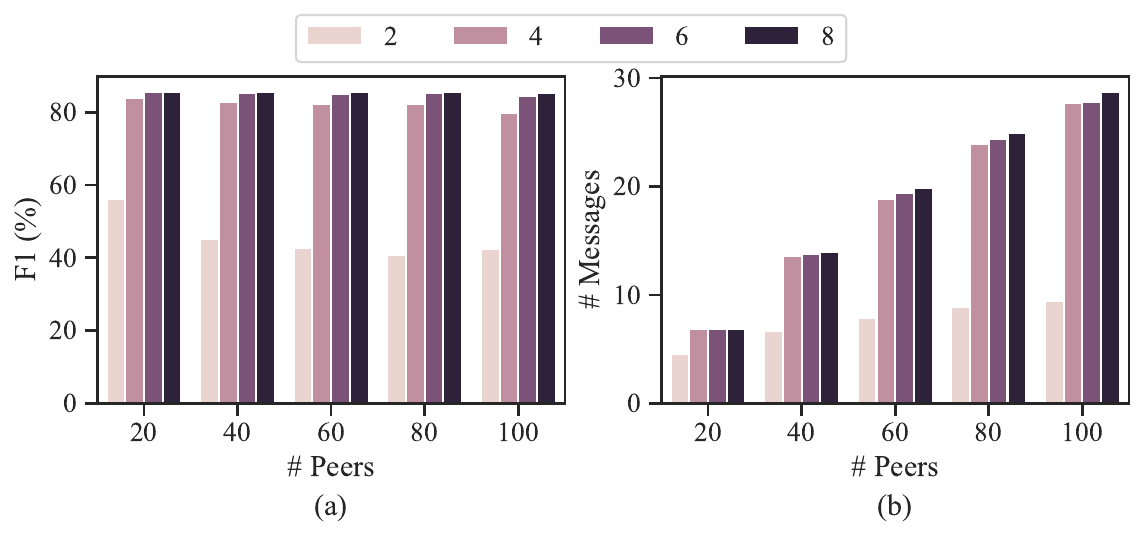}
    \caption{Influence of the neighbor selection count, $k$, on performance (a) and communication cost (b) in DRAG-TARW. $k$ represents the number of query neighbors considered for forwarding the query in each hop ($k = 2, 4, 6, 8$).}
    \label{fig:sensitivity_query_neighbor}
\end{figure}

The experimental results, presented in Table~\ref{table:drag_baseline}, provide a comparison of the proposed DRAG framework with TARW against several baselines: NoRAG (no retrieval), Centralized RAG (CRAG), DRAG with Random Walk (DRAG-RW), and DRAG with Flooding (DRAG-FL). A key finding is that DRAG-TARW achieves performance comparable to CRAG across all three datasets (MMLU, Medical Extended, and News) in terms of Exact Match (EM), F1 Score, Precision, and Recall, demonstrating that the distributed approach maintains accuracy without a centralized knowledge base. For instance, on the MMLU dataset, DRAG-TARW achieves an EM of $83.90\%$ compared to CRAG's $85.73\%$, and an F1 score of $83.92\%$ compared to CRAG's $85.75\%$. Critically, DRAG-TARW significantly reduces communication overhead compared to DRAG-FL, as evidenced by the lower number of messages required for knowledge retrieval. On MMLU, DRAG-TARW uses an average of $6.87$ messages compared to DRAG-FL's $10.91$, and on News, DRAG-TARW uses $7.82$ messages compared to DRAG-FL's $10.99$. This highlights the efficiency of the TARW algorithm in identifying relevant peers and routing queries effectively within a distributed environment. DRAG-RW performs substantially worse than all other RAG variants, particularly in terms of Hit Rate, indicating that undirected exploration of the network struggles to locate relevant knowledge. For example, the Hit Rate of DRAG-RW on MMLU is only $23.46\%$, compared to $98.11\%$ for DRAG-TARW. While DRAG-TARW achieves lower message counts than DRAG-FL, there is a slight trade-off in accuracy, with DRAG-FL generally exhibiting marginally higher EM, F1, Precision, and Recall scores. The choice between DRAG-TARW and DRAG-FL depends on the specific application context and the relative importance of accuracy versus communication efficiency. Finally, the NoRAG baseline clearly demonstrates the necessity of retrieval augmentation, as the LLM's performance without external knowledge is significantly lower than any RAG configuration, highlighting the benefits of integrating external knowledge to enhance accuracy and completeness. For example, on the MMLU dataset, NoRAG achieves an F1 score of only $17.41\%$, demonstrating the substantial improvement provided by retrieval-augmented generation.

The experimental results presented in Figure~\ref{fig:f1_scheme_datasets} demonstrate the performance of the DRAG framework against Centralized RAG and its variants with incomplete knowledge bases. DRAG consistently achieves F1 scores comparable to CRAG when a complete knowledge base is available. For instance, on the Medical Extended dataset, DRAG achieves an F1 score of $90.58\%$ compared to CRAG's $94.46\%$. Notably, DRAG significantly outperforms CRAG when the centralized knowledge base is incomplete, either due to a reduction in the number of samples (CRAG-0.7S, CRAG-0.5S) or a reduction in the number of topics covered (CRAG-0.7T, CRAG-0.5T). On the MMLU dataset, when only $70\%$ of the samples are available (CRAG-0.7S), CRAG's F1 score drops to $61.54\%$, significantly lower than DRAG's $83.92\%$. This resilience to incomplete knowledge highlights a key advantage of DRAG: its ability to leverage a distributed knowledge base to compensate for gaps in any single node's information. This advantage is especially noticeable in real-world situations when it is infeasible to maintain a complete and up-to-date centralized knowledge base, particularly in light of privacy concerns.

\subsection{Sensitivity Analysis}

Figure~\ref{fig:f1_anm_peer} presents a sensitivity analysis of the DRAG framework to variations in network size, examining the impact on both F1 score and message overhead. As the number of peers increases from $20$ to $100$, DRAG-TARW experiences a modest decrease in F1 score. Specifically, on the MMLU dataset, the F1 score decreases from $83.92\%$ to $79.72\%$, while on the Medical Extended dataset, it decreases from $90.58\%$ to $86.66\%$. Despite this slight reduction in F1 score, DRAG-TARW consistently outperforms DRAG-RW, which exhibits a far more significant drop as the network grows. For example, on MMLU, the F1 score of DRAG-RW plummets from $19.97\%$ at $20$ peers to just $3.57\%$ at $100$ peers. Simultaneously, while the average number of messages for both DRAG-TARW and DRAG-FL increases with network size, DRAG-TARW consistently maintains a significantly lower message overhead than DRAG-FL. Quantitatively, at $100$ peers on MMLU, DRAG-TARW requires only $27.67$ messages compared to DRAG-FL's $53.34$, and on News, it requires $23.71$ messages versus $48.83$ for DRAG-FL, representing approximately a $50\%$ reduction in message overhead. This represents a substantial advantage in resource-constrained edge environments. In addition, these results suggest a potential trade-off between network size and individual peer performance in DRAG, and emphasize the need for further optimization of the knowledge searching algorithm for larger networks.

Figure~\ref{fig:anm_nq} examines the relationship between the number of queries processed and the average number of messages required by DRAG-TARW on the MMLU dataset, revealing a beneficial convergence effect. The data demonstrates that as the number of queries increases, the average number of messages per query gradually decreases, eventually stabilizing at a consistent level. For example, in a network with $100$ peers, the average number of messages starts at $34.73$ for the first $500$ queries but converges to $27.67$ after processing $10000$ queries, representing a significant reduction of over $20\%$. This convergence is observed across all network sizes ($20$ to $100$ peers), indicating that DRAG-TARW becomes increasingly efficient as peers progressively cache more topics and accumulate knowledge about their neighbors' expertise in specific topics over time. This suggests that DRAG-TARW is particularly well-suited for scenarios with sustained query loads or prior knowledge of neighbors' expertise.

Figure~\ref{fig:sensitivity_peer_attach} presents a sensitivity analysis on the impact of peer connectivity within the DRAG-TARW framework on the MMLU dataset. The results reveal that increasing peer connectivity does not consistently lead to either improved F1 scores or reduced message overhead. Despite varying the number of connections per peer from $2$ to $8$, the F1 score remains relatively stable, fluctuating between approximately $80\%$ and $83\%$. Moreover, while one might expect increased connectivity to facilitate more direct routing and fewer messages, the average number of messages could increase with the number of attached edges, particularly in larger networks. For instance, in a network of $100$ peers, increasing the connectivity from $2$ to $8$ results in an increase in the average number of messages from $26.64$ to $29.37$. This lack of consistent improvement in both F1 and message count likely stems from the inherent randomness in knowledge distribution across the peer-to-peer network and the complex interplay with the underlying network topology. With randomly distributed knowledge, more connections do not guarantee access to more relevant knowledge, and the specific arrangement of connections can influence routing paths in unpredictable ways. These results highlight the challenges of optimizing network topology in DRAG systems and underscore the importance of considering the interplay between knowledge distribution, network structure, and routing algorithm design.

Figure~\ref{fig:sensitivity_llm_datasets} presents a sensitivity analysis evaluating the impact of different LLMs on the performance and communication cost of DRAG-TARW across the MMLU, Medical Extended, and News datasets. The results underscore that the choice of LLM significantly influences the overall performance of DRAG-TARW, as evidenced by the variability in F1 scores across different models for a given dataset. For instance, on the MMLU dataset, Llama 3.2 3B achieves an F1 score of $83.92\%$, while Gemma 2 2B scores $77.13\%$ and Qwen2.5 3B achieves $82.86\%$. This performance variation can be attributed, in part, to the LLM's ability to effectively extract the underlying topic from the query, which directly impacts the efficiency of the knowledge search process within DRAG-TARW. However, the data also reveals that a higher F1 score does not necessarily correlate with a higher average number of messages. For example, on the News dataset, Llama 3.2 3B, despite achieving a moderate F1 score ($74.73\%$), requires $7.82$ messages, while Gemma 2 2B, with a slightly higher F1 score ($76.12\%$), only requires $6.90$ messages. This decoupling of F1 score and message overhead is likely due to a combination of factors, including the inherent randomness in the network topology and the stochastic nature of LLM generation itself. 

Figure~\ref{fig:sensitivity_query_neighbor} presents a sensitivity analysis examining the impact of the number of query neighbors ($k$) on the performance and communication cost of DRAG-TARW. The results demonstrate a clear trade-off between F1 score and message overhead. Increasing the number of query neighbors initially leads to a significant improvement in F1 score, particularly when transitioning from $k=2$ to $k=4$. For instance, in a network of $20$ peers, increasing $k$ from $2$ to $4$ results in a jump in F1 score from $56.27\%$ to $83.92\%$. However, beyond $k=4$, further increases in the number of query neighbors yield only marginal improvements in F1 score, while substantially increasing the average number of messages. In a network of $100$ peers, the F1 score increases from $79.72\%$ at $k=4$ to only $85.33\%$ at $k=8$, while the average number of messages increases from $27.67$ to $28.71$. This suggests that while considering more neighbors initially helps to broaden the search and improve retrieval accuracy, the benefits diminish as $k$ increases, and the additional communication overhead outweighs the gains. These findings underscore the importance of carefully selecting the number of query neighbors to optimize the balance between performance and efficiency in DRAG-TARW.

\section{Conclusion}
\label{sec:conclusion}

This paper introduced Distributed Retrieval-Augmented Generation, a novel framework that enhances data privacy and scalability by eliminating the need for a centralized knowledge base and empowering data owners with control over their information. DRAG leverages a Topic-Aware Random Walk algorithm to efficiently discover relevant peers and retrieve knowledge within a P2P network. Extensive experiments across diverse datasets and LLMs demonstrated that DRAG with TARW achieves near-centralized RAG performance while significantly reducing communication overhead compared to flooding-based approaches. The results highlight DRAG as a promising scheme in privacy-sensitive and resource-constrained edge environments. Future work will focus on further refining the knowledge searching algorithm to improve efficiency and scalability in very large and dynamic networks.

\bibliographystyle{IEEEtran}
\bibliography{ref}

\end{document}